\documentclass[letter,longauth,traditabstract]{aa}  
\usepackage{graphicx}
\usepackage{txfonts}
\usepackage[authoryear]{natbib}
\bibpunct{(}{)}{;}{a}{}{,}

\newcommand{\den}{${n}_{\mathrm{H}_2}$}
\newcommand{\ncolone}{$N(^{12}$CO)}
\newcommand{\ncoltwo}{$N(^{13}$CO)}
\newcommand{\tkin}{$T_{\mathrm{kin}}$}
\newcommand{\cmtwo}{cm$^{-2}$}
\newcommand{\cmthree}{cm$^{-3}$}
\newcommand{\coone}{$^{12}$CO}
\newcommand{\cotwo}{$^{13}$CO}
\newcommand{\msun}{M$_{\sun}$}

\begin{document}

\title{Probing the molecular interstellar medium of \object{M82} with Herschel-SPIRE
spectroscopy\thanks{\textit{Herschel} is an ESA space observatory with science instruments
provided by European-led Principal Investigator consortia and with important participation
from NASA.}}

\titlerunning{Probing M82 molecular ISM with SPIRE FTS}

\author{P. Panuzzo\inst{\ref{inst11}},
N. Rangwala\inst{\ref{inst13}},
A. Rykala\inst{\ref{inst2}},
K. G. Isaak\inst{\ref{inst2},\ref{inst23}},
J. Glenn\inst{\ref{inst13}},
C. D. Wilson\inst{\ref{inst20}},
R. Auld\inst{\ref{inst2}},
M. Baes\inst{\ref{inst5}},
M. J. Barlow\inst{\ref{inst6}},
G. J. Bendo\inst{\ref{inst3}},
J. J. Bock\inst{\ref{inst7}},
A. Boselli\inst{\ref{inst1}},
M. Bradford\inst{\ref{inst7}},
V. Buat\inst{\ref{inst1}},
N. Castro-Rodr{\'\i}guez\inst{\ref{inst8}},
P. Chanial\inst{\ref{inst11}},
S. Charlot\inst{\ref{inst9}},
L. Ciesla\inst{\ref{inst1}},
D. L. Clements\inst{\ref{inst3}},
A. Cooray\inst{\ref{inst25}},
D. Cormier\inst{\ref{inst11}},
L. Cortese\inst{\ref{inst2}},
J. I. Davies\inst{\ref{inst2}},
E. Dwek\inst{\ref{inst10}},
S. A. Eales\inst{\ref{inst2}},
D. Elbaz\inst{\ref{inst11}},
T. Fulton\inst{\ref{inst28}},
M. Galametz\inst{\ref{inst11}},
F. Galliano\inst{\ref{inst11}},
W. K. Gear\inst{\ref{inst2}},
H. L. Gomez\inst{\ref{inst2}},
M. Griffin\inst{\ref{inst2}},
S. Hony\inst{\ref{inst11}},
L. R. Levenson\inst{\ref{inst7}},
N. Lu\inst{\ref{inst7}},
S. Madden\inst{\ref{inst11}},
B. O'Halloran\inst{\ref{inst3}},
K. Okumura\inst{\ref{inst11}},
S. Oliver\inst{\ref{inst14}},
M. J. Page\inst{\ref{inst15}},
A. Papageorgiou\inst{\ref{inst2}},
T. J. Parkin\inst{\ref{inst20}},
I. P{\'e}rez-Fournon\inst{\ref{inst8}},
M. Pohlen\inst{\ref{inst2}},
E. T. Polehampton\inst{\ref{inst26},\ref{inst27}},
E. E. Rigby\inst{\ref{inst4}},
H. Roussel\inst{\ref{inst9}},
N. Sacchi\inst{\ref{inst17}},
M. Sauvage\inst{\ref{inst11}},
B. Schulz\inst{\ref{inst16}},
M. R. P. Schirm\inst{\ref{inst20}},
M. W. L. Smith\inst{\ref{inst2}},
L. Spinoglio\inst{\ref{inst17}},
J. A. Stevens\inst{\ref{inst18}},
S. Srinivasan\inst{\ref{inst9}},
M. Symeonidis\inst{\ref{inst15}},
B. Swinyard\inst{\ref{inst26}},
M. Trichas\inst{\ref{inst3}},
M. Vaccari\inst{\ref{inst19}},
L. Vigroux\inst{\ref{inst9}},
H. Wozniak\inst{\ref{inst21}},
G. S. Wright\inst{\ref{inst24}},
W. W. Zeilinger\inst{\ref{inst22}}
}

\institute{	
CEA, Laboratoire AIM, Irfu/SAp, Orme des Merisiers, F-91191
Gif-sur-Yvette, France\label{inst11}
\and
Dept. of Astrophysical \& Planetary Sciences, CASA CB-389,  
University of Colorado, Boulder, CO 80309, USA\label{inst13}
\and
School of Physics \& Astronomy, Cardiff University, Queens  
Buildings The Parade, Cardiff CF24 3AA, UK\label{inst2}
\and
ESA Astrophysics Missions Division, ESTEC, PO Box 299, 2200 AG
Noordwijk, The Netherlands\label{inst23} 
\and
Dept. of Physics \& Astronomy, McMaster University, Hamilton,  
Ontario, L8S 4M1, Canada\label{inst20}
\and
Sterrenkundig Observatorium, Universiteit Gent, Krijgslaan 281 S9,  
B-9000 Gent, Belgium\label{inst5}
\and
Dept. of Physics \& Astronomy, University College London,  
Gower Street, London WC1E 6BT, UK\label{inst6}
\and
Astrophysics Group, Imperial College, Blackett Laboratory, Prince  
Consort Road, London SW7 2AZ, UK\label{inst3}
\and
JPL, Pasadena, CA 91109, United States;  
Dept. of Astronomy, California Institute of Technology, Pasadena,  
CA 91125, USA\label{inst7}
\and
Laboratoire d'Astrophysique de Marseille, UMR6110 CNRS, 38 rue F.  
Joliot-Curie, F-13388 Marseille, France\label{inst1}
\and
Instituto de Astrof{\'\i}sica de Canarias (IAC) and Departamento de
Astrof{\'\i}sica, Universidad de La Laguna (ULL), La Laguna, Tenerife, 
Spain\label{inst8}
\and
Institut d'Astrophysique de Paris, UMR7095 CNRS, Universit\'e Pierre  
\& Marie Curie, 98 bis Boulevard Arago, F-75014 Paris, France\label{inst9}
\and
Dept. of Physics \& Astronomy, University of California, Irvine,
CA 92697, USA\label{inst25}
\and	
Observational Cosmology Lab, Code 665, NASA Goddard Space Flight   
Center Greenbelt, MD 20771, USA\label{inst10}
\and
Blue Sky Spectroscopy, Alberta, T1J 1B1, Canada\label{inst28}
\and
Dept. of Physics and Astronomy, University of  
Sussex, Brighton, BN1 9QH, UK\label{inst14}
\and
Mullard Space Science Laboratory, University College London,  
Holmbury St Mary, Dorking, Surrey RH5 6NT, UK\label{inst15}
\and
Space Science and Technology Department, Rutherford Appleton Laboratory, 
Oxfordshire, OX11 0QX, UK\label{inst26}
\and
Institute for Space Imaging Science, University of Lethbridge, Lethbridge, 
Alberta, T1K 3M4, Canada\label{inst27}
\and
School of Physics \& Astronomy, University of Nottingham, University  
Park, Nottingham NG7 2RD, UK\label{inst4}
\and
Istituto di Fisica dello Spazio Interplanetario, INAF, Via del Fosso  
del Cavaliere 100, I-00133 Roma, Italy\label{inst17}
\and
IPAC, California Institute of  
Technology, Mail Code 100-22, 770 South Wilson Av, Pasadena, CA 91125,  
USA\label{inst16}
\and
Centre for Astrophysics Research, Science \& Technology Research  
Centre, University of Hertfordshire, College Lane, Herts AL10 9AB, UK\label{inst18}
\and
University of Padova, Dept. of Astronomy, Vicolo Osservatorio  
3, I-35122 Padova, Italy\label{inst19}
\and
Observatoire Astronomique de Strasbourg, UMR 7550 Universit\'e de  
Strasbourg - CNRS, 11, rue de l'Universit\'e, F-67000 Strasbourg, France\label{inst21}
\and
UK Astronomy Technology Center, Royal Observatory Edinburgh,
Edinburgh, EH9 3HJ, UK\label{inst24} 
\and
Institut f\"ur Astronomie, Universit\"at Wien, T\"urkenschanzstr. 17,  
A-1180 Wien, Austria\label{inst22}
}

\authorrunning{Panuzzo et al.}

\date{Received ????; accepted ????}

% \abstract{}{}{}{}{} 
% 5 {} token are mandatory
 
\abstract{We present the observations of the starburst galaxy \object{M82} taken with 
the \textit{Herschel} SPIRE Fourier Transform Spectrometer. The spectrum (194--671 $\mu$m)
shows a prominent CO rotational ladder from $J$ = 4--3 to 13--12 emitted by the central region
of \object{M82}. The fundamental properties of the gas are well constrained by the high $J$
lines observed for the first time. Radiative transfer modeling of these high-S/N $^{12}$CO and
$^{13}$CO lines strongly indicates a very warm molecular gas component at $\sim 500$~K and
pressure of $\sim 3 \times 10^{6}$~K~cm$^{-3}$,  in good agreement with the H$_{2}$ rotational lines
measurements from \textit{Spitzer} and \textit{ISO}. We suggest that this warm gas is heated
by dissipation of turbulence in the interstellar medium (ISM) rather than X-rays or UV flux 
from the straburst. This paper illustrates the promise of the SPIRE FTS for the study of
the ISM of nearby galaxies.}


\maketitle
%
%________________________________________________________________

\section{Introduction}
Starburst galaxies provide us with the opportunity to study star
formation and its effect on the interstellar medium (ISM) in 
extreme environments. These galaxies combine large central gas concentrations
and high ionizing radiation fields, resulting in bright molecular, neutral
and ionized gas emission lines.

At a distance of 3.9 Mpc \citep{sakai99}, \object{M82} is the most
well-studied starburst galaxy in the local universe, and it is
widely used as a starburst prototype in cosmological studies.
Its infrared luminosity \citep[$5.6\times 10^{10}$ L$_\odot$,][]{sanders03}
corresponds to a star-formation rate of 9.8 M$_\odot$ yr$^{-1}$, which has
almost certainly been enhanced by its interaction with 
\object{M81} and \object{NGC 3077} \citep{yun93}. 
With a reported molecular gas content of $1.3 \times 10^9$ M$_\odot$ \citep{walter02}, 
its bright emission lines of CO and other molecules allow us to study 
its ISM in great detail \citep{shen95,walter02,ward03}.
 
Far-infrared fine structure lines were used to constrain the physical properties
of the ionized gas and photo-dissociation regions (PDRs) in \object{M82}. 
\citet{colbert99} found that the ionized gas emission can be reproduced
with a 3--5 Myr old instantaneous starburst 
and a gas density of 250 cm$^{-3}$, while the PDR
component is best fit with a density of 2\,000 cm$^{-3}$,
in pressure equilibrium with the ionized phase.

Stellar evolution and photoionization models \citep{forster03} indicate 
a series of a few, Myr-duration starbursts with a peak of activity 10
Myr ago in the central regions, and 5 Myr ago in the circumnuclear
ring. Models of the PDR and molecular emission as a set of
non-interacting hot bubbles driving spherical shells of swept-up gas
into a surrounding uniform medium also
predict a starburst age of 5--10 Myr, 
but fail to match the observed far-infrared luminosity \citep{yao09}.

The strengths of the CO lines place fundamental constraints on the physical properties 
of the molecular gas. \cite{tilanus91} fitted
\coone\ and \cotwo\ lines from the central starburst up to $J$ = 3--2 
with a single-component model with temperatures of 30--55 K and densities of 
3--7 $\times 10^3$ cm$^{-3}$.
\citet{wild92} used lines up to the CO $J$ = 6--5 transition to refine these parameters to
40--50 K and $\sim 10^4$ cm$^{-3}$, while HCN and HCO$^+$ lines suggested densities greater than $3\times 10^5$
cm$^{-3}$ are present.
\citet{petitpas00} showed evidence for a temperature or density gradient across the starburst region. 
\citet{weiss05} showed that CO emission up to $J$ = 3--2 is dominated by more extended regions while
higher $J$ transitions originate in the central disk.

In this paper, we present observations of \object{M82} with
\textit{Herschel} \citep{pilbratt10} using the 
SPIRE Fourier Transform Spectrometer (FTS) \citep{griffin10}, which
measures the complete far-infrared spectrum from
194 to 671 $\mu$m. This spectral region is particularly
interesting for probing the peak of the CO spectral line energy distribution 
(SLED) in gas-rich galaxies.
The wealth of lines across a continuous spectral region allows for
unprecedented precision in modeling the physical and chemical
properties of the molecular ISM.
Here, we focus on the measurement and analysis of the CO rotational
transitions from the central starburst in \object{M82}.

%%%%%%%%%%%%%%%%%%%%%%%%%%%%%%%%%%%%%%%%%%%%%%%%%%%%%%%%%%%%%%%%%%%%%%%%%%%%%

\section{Observations and data reduction}\label{sec_observ}

\begin{figure}
\centering
\includegraphics[width=0.45\textwidth]{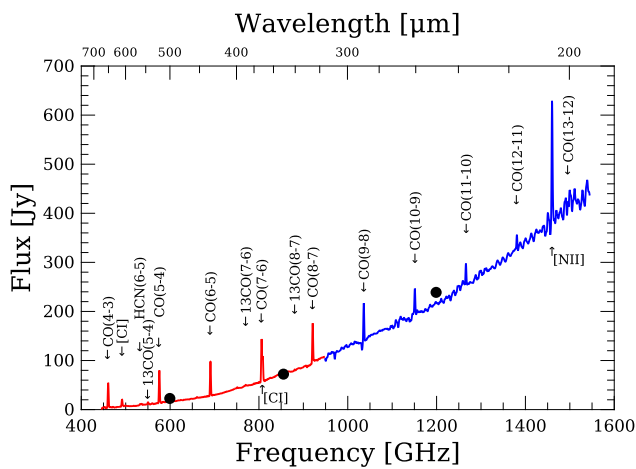}
\caption{Apodized spectrum of \object{M82} corresponding to a 43\farcs4 beam, where red and blue lines represent
data from the long- and the short-wavelength FTS bands respectively. Filled circles show 
SPIRE fluxes measured in the same beam.}
\label{fig_m82_spectrum}
\end{figure}

%%%%%%%%%%%%%%%%%%%%%%%%%

The galaxy \object{M82} was observed by the SPIRE FTS in the high spectral resolution (FWHM=0.048 cm$^{-1}$), 
point-source mode, on 2009 September 21 as a performance verification target. 
The total integration time was 1\,332 seconds.
The data were processed and calibrated as described in \citet{swinyard10}. 
Only data from the central detectors in the two FTS bands are presented here.

The beam size of the FTS bolometers varies with wavelength across the 
individual bands \citep[see][]{swinyard10}, and the spatial extent of the
\object{M82} central starburst is comparable to the beam size 
(mean FWHM $\sim$19\arcsec\ and 35\arcsec\ for the short- and the long-wavelength bands respectively).
For a proper comparison with models, the spectrum must be scaled
appropriately to a single beam size by a source-beam coupling factor
($\eta_c(\nu)$). This factor was obtained by
convolving the \object{M82} SPIRE photometer map at 250~$\mu$m
\citep{roussel10}, which has a beam FWHM of 18\farcs1
\citep{griffin10}, with appropriate Gaussian profiles to reproduce the
light distribution as seen by FTS bolometers at different beam
sizes. 
The value of $\eta_c(\nu)$ is then given by the ratio of the beam-integrated flux density of the map convolved to the
beam size corresponding to the given frequency ($\nu$) to the
beam-integrated flux density of the map with the largest beam size (43\farcs4); its values goes from 1 to 0.42.
This implicitly assumes that the dust and CO emission distributions within the
beam are the same at all frequencies.

We opted to use the extended-source calibrated\footnote{ Extended-source flux calibration is derived
from telescope emission measurements, while the point-source flux calibration is based on
observations of known astronomical point sources.} spectrum because the 
point-source calibration was more noisy and suffered from significant
uncertainties below 600 GHz. 
We found, however, that the extended-source calibrated
spectrum corrected for source-beam coupling is around a factor of 2 fainter
than photometry for the same beam. We thus scaled the spectrum
to match the photometry in the three bands 
by applying a single constant scaling factor 
for the short-wavelength band and a factor with a linear dependence on frequency for the
long-wavelength band.
The resulting spectrum is shown in Fig. \ref{fig_m82_spectrum} 
\citep[for clarity, we show the spectrum apodized using the
extended Norton-Beer function 1.5 from][]{naylor07}; we note that
the short- and long-wavelength bands match very smoothly.

Line fluxes were recovered from the calibrated unapodized spectrum 
using a custom-written tool. It first subtracts the underlying continuum using a 
grey-body fit, then it removes any remaining large-scale ripples using a polynomial function. 
Emission lines were extracted by fitting sinc-convolved Gaussian line profiles.
The strongest line is fitted first and then subtracted, with the process repeated until 
no line is found above a pre-set discrimination level. 
The integrated line fluxes were obtained by calculating the area under the fit.
Table \ref{table_lines} lists the line fluxes and their 1$\sigma$ uncertainties derived from
the fitting procedure.
In addition to the reported uncertainties we should include the following contributions:
(i) the uncertainty in the estimation of the source-beam coupling factor due
to the uncertainty in the beam profile and the assumption of identical
distributions for dust and CO emission, and (ii) the uncertainty involved in the scaling the
spectrum to match the photometric data, and in the measurement of photometric data.
We conservatively suggest an uncertainty of
$\sim$30\% for the line fluxes due to the above factors.

\begin{figure}
\centering
\includegraphics[width=0.35\textwidth,bb=78 371 365 565,clip]{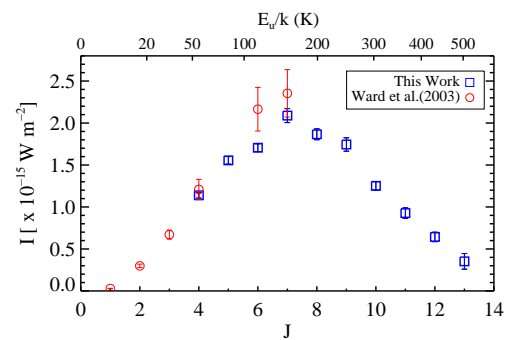}
\caption{$^{12}$CO SLED for \object{M82} for a 43\farcs4 beam as measured in this work (open squares) 
with 1$\sigma$ statistical error bars. \citet{ward03} ground based data are shown by open circles.}
\label{fig_sled}
\end{figure}

Fig. \ref{fig_sled} shows the $^{12}$CO SLED, which peaks at the $J$ = 7--6 line.
In the same plot, we draw ground-based data compiled by \citet{ward03} (W03 hereafter).
These data were measured with a smaller beam size,
but were given for two observed lobes. In the plot we used the sum of the fluxes measured in the two lobes
(which have a small overlap, but fit within the 43\farcs4 beam).

\begin{table}
\caption{\label{tab_lines}Measured fluxes of detected emission lines. Errors are 1$\sigma$ only from line fitting procedure, not including other uncertainties (see text). The $^{13}$CO $J$=6--5 line is missing due to fringing.}
\label{table_lines}
\centering
\begin{tabular}{l r r r}
\hline\hline                 % inserts double horizontal lines
Transition name & Frequency & \multicolumn{2}{c}{Flux}   \\    % table heading
     & (rest, GHz) & (10$^3$ Jy km s$^{-1}$) & (10$^{-16}$ W m$^{-2}$) \\    % table heading
\hline                        % inserts single horizontal line
$^{12}$CO $J$ = 4--3   &   461.041 &  74.1 $\pm$ 2.2 & 11.32 $\pm$ 0.33 \\
$^{12}$CO $J$ = 5--4   &   576.268 &  80.9 $\pm$ 2.3 & 15.53 $\pm$ 0.45 \\
$^{12}$CO $J$ = 6--5   &   691.473 &  74.0 $\pm$ 2.0 & 17.04 $\pm$ 0.46 \\
$^{12}$CO $J$ = 7--6   &   806.652 &  77.7 $\pm$ 3.1 & 20.89 $\pm$ 0.84 \\
$^{12}$CO $J$ = 8--7   &   921.800 &  60.7 $\pm$ 2.1 & 18.64 $\pm$ 0.65 \\
$^{12}$CO $J$ = 9--8   &  1036.912 &  50.5 $\pm$ 2.3 & 17.44 $\pm$ 0.79 \\
$^{12}$CO $J$ = 10--9  &  1151.985 &  32.6 $\pm$ 1.3 & 12.51 $\pm$ 0.50 \\
$^{12}$CO $J$ = 11--10 &  1267.014 &  21.9 $\pm$ 1.5 &  9.28 $\pm$ 0.63 \\
$^{12}$CO $J$ = 12--11 &  1381.995 &  14.0 $\pm$ 1.2 &  6.44 $\pm$ 0.57 \\
$^{12}$CO $J$ = 13--12 &  1496.922 &   7.1 $\pm$ 1.9 &  3.53 $\pm$ 0.93 \\
$^{13}$CO $J$ = 5--4   &   550.926 &   5.3 $\pm$ 0.7 &  0.98 $\pm$ 0.12 \\
$^{13}$CO $J$ = 7--6   &   771.184 &   3.2 $\pm$ 0.6 &  0.81 $\pm$ 0.16 \\
$^{13}$CO $J$ = 8--7   &   881.273 &   2.3 $\pm$ 0.7 &  0.68 $\pm$ 0.22 \\
HCN $J$ = 6--5         &   531.716 &   2.9 $\pm$ 0.7 &  0.52 $\pm$ 0.12 \\
$[$\ion{C}{I}$]~^3P_1\rightarrow\ ^3P_0$  &  492.161 &  20.6 $\pm$ 1.6 & 3.38 $\pm$ 0.26\\
$[$\ion{C}{I}$]~^3P_2\rightarrow\ ^3P_1$  &  809.342 &  43.2 $\pm$ 0.9 &11.66 $\pm$ 0.25\\
$[$\ion{N}{II}$]~^3P_1\rightarrow\ ^3P_0$ & 1462.000 & 124.1 $\pm$ 5.8 &60.51 $\pm$ 2.85\\      
\hline                                   %inserts single line
\end{tabular}
\end{table}

%%%%%%%%%%%%%%%%%%%%%%%%%%%%%%%%%%%%%%%%%%%%%%%%%%%%%%%%%%%%%%%%%%%%%%%%%%%

\begin{figure*}
\centering
\includegraphics[angle=90,width=0.75\textwidth,bb=62 82 219 732,clip]{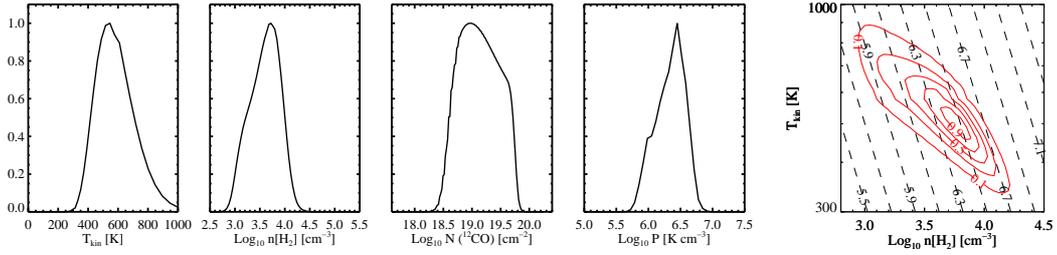}
\caption{Left Panel: Likelihood distributions of kinetic temperature, density, CO column density and pressure.
Right Panel: Likelihood contour plot of temperature and density. Dashed lines show constant pressure (Log$_{10}$ $P$ (K cm$^{-3}$)) relations.}
\label{like}
\end{figure*}

\section{Radiative transfer modeling}

We used RADEX \citep{vandertak07}, a non-LTE code that computes the intensities of molecular 
lines by iteratively solving for statistical equilibrium using an escape-probability formalism
assuming a uniform expanding sphere, to model the CO line intensities. 
The main inputs to RADEX are the gas density (\den), the kinetic temperature (\tkin), and the CO column density
per unit line width (\ncolone/$\Delta \mathrm{v}$). 
We ran the code for a large parameter grid in \tkin\ (20--1\,000 K), \den\ ($10^{2}$--$10^{6}$ 
\cmthree), \ncolone\ ($10^{15}$--$10^{18}$ \cmtwo), and \ncoltwo\ ($10^{13}$--$10^{17}$\cmtwo).
From this grid of models, we generated likelihood distributions by adapting the method described
in W03, for \tkin, \den, \ncolone, \ncoltwo, and pressure by comparing the RADEX and observed line fluxes. 

To avoid any non-physical situation we applied two priors in this analysis following W03.
The first one limits the $^{12}$CO column density in a way
that the total mass of the molecular gas producing the CO lines cannot exceed the dynamical mass of the
galaxy according to the following relation:
\begin{equation}
N(^{12}\mathrm{CO}) < \frac{M_\mathrm{{dyn}}x_\mathrm{CO}}{\mu m_\mathrm{H_{2}}}\frac{1}{\pi R_\mathrm{d}^{2}} 
= 2.3 \times 10^{20}\, \mathrm{cm}^{-2} ~,
\end{equation}
where the dynamical mass of the disk $M_\mathrm{dyn} = 2.4 \times 10^{9}$ \msun, the radius of the disk 
$R_\mathrm{d} = 250\, \mathrm{pc}$, the abundance of CO relative to H$_{2}$, $x_{\mathrm{CO}} = 3 \times 10^{-4}$
(W03), and $\mu = 1.4$ is the mean molecular weight in units of $m_\mathrm{H_{2}}$.
The second prior limits the column length to be less than that of the entire molecular region according to
$\frac{N(\mathrm{CO})}{n(\mathrm{H_{2}})x_{\mathrm{CO}}}  < 1.54 \times 10^{21}\, \mathrm{cm}$. In this
analysis we used all the $^{12}$CO and $^{13}$CO lines in Table \ref{table_lines} along with their 1$\sigma$
statistical errors. It
was necessary to add 20\% and 10\% uncertainties for the CO $J \le$  8--7 and $J>$  8--7 lines, respectively,
to avoid un-physically
narrow and noisy distributions (consistent with the additional 30\% line flux
uncertainty estimate in Sect. \ref{sec_observ}). 
The resulting distributions are shown in Fig. \ref{like} for each variable, marginalizing over the other variables.
The modeling only depends on the relative line fluxes, therefore the results will not be affected by the uncertainties in the absolute flux calibration.

\begin{figure}
\center
\includegraphics[width=0.3\textwidth,bb=75 376 431 623]{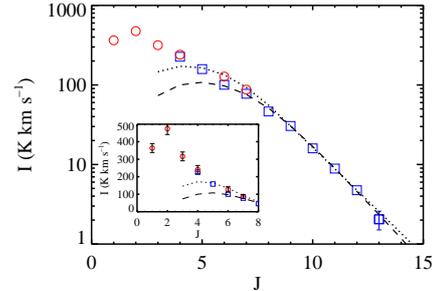}
\caption{Comparing the highest likelihood model (dotted line) with our CO line intensities. The model shown by a dashed
line was obtained by using only $\mathrm{J} \geq 7-6$ CO lines.  The W03 data are shown by open circles.
The inset highlights the deviations from models at the lower J end.}\label{model}
\end{figure}

%%%%%%%%%%%%%%%%%%%%%%%%%%%%%%%%%%%%%%%%%%%%%

\section{Results and discussion}

We found that the highest likelihood model (dotted line in Fig. \ref{model}) provides a good fit to our data
(open squares), in particular for the higher $J$ lines ($J \geq $ 7--6). The likelihood contour plot of
temperature and density
in Fig. \ref{like} (last panel) strongly indicates that the observed emission is coming from very warm gas
with a kinetic temperature of $\sim$ 540~K and a pressure of $\sim$~$3 \times 10^{6}$~K~cm$^{-3}$.
The detailed physical characteristics of the warm gas are listed in Table \ref{besttable}, which are obtained from the likelihood distributions shown in Fig. \ref{like}. 
%The detailed physical characteristics of the warm gas, obtained from the likelihood distributions of Fig. \ref{like}, are listed in Table \ref{besttable}.

The ISM of this galaxy has been well-studied using ground-based observations -- in particular the lowest-lying CO rotational
lines that provide constraints on the physical state of the cold molecular gas. Several studies from ground-based CO
observations, including W03, have identified cold gas at $\sim$30~K. We show W03 data (open circles) over-plotted in
Figure \ref{model}. From $J \leq $ 7--6 lines, W03 deduced the presence of a warm component, in agreement with our finding, but with different temperature, mass and density. 
Having observations up to $J$ = 13--12 enables a much better constraint to be placed
on these parameters than is possible from the lower $J$ lines.
A colder gas component is also consistent with the deviation of our lower $J$ lines, especially
CO $J$ = 4--3, from our highest-likelihood model. If we use only the higher-$J$ lines ($J \geq $ 7--6)
in our likelihood analysis, we get a model in better agreement with these lines while underpredicting
the lower-$J$ lines, supporting the hypothesis of a contribution of the colder component to those lines.

Assuming $x_{\mathrm{CO}}$ of $3 \times 10^{-4}$ and an intrinsic line width of 180 km s$^{-1}$ (W03),
and using our beam-averaged CO column density we find the mass of warm gas to be $1.2 \times 10^{7}$ M$_{\sun}$
within a beam area of about 2140 square arcseconds, likely covering most of the molecular emission from the galaxy.
Using the mass of the cold gas from W03 we find a ratio of cold gas to the SPIRE-observed warm gas mass of $\sim 3$.
The best-fitting model predicts the optical depth for the CO lines, which peaks at a value of 1.7 for $J$ = 6--5, 
and then drops to approximately $10^{-2}$ for $J$ = 13--12.

%%%%%%%%%%%%%

Mid-IR H$_{2}$ rotational lines are optically-thin and easily thermalized, so they provide an
independent constraint on the mass of warm gas.  Several transitions have been studied with \textit{ISO}
\citep{rigopoulou02} and \textit{Spitzer} \citep{beirao08}. 
Both studies agree that the S(1) to S(2) line ratio suggests $T \sim 500$~K (assuming an ortho-to-para ratio of 3),
in excellent agreement with our temperature. Using the \textit{Spitzer} measurement of S(1) line flux corrected for our
larger beam we calculate a mass of $\sim 2 \times 10^{6}\, \mathrm{M}_{\sun}$. Given the uncertainty on $x_{\mathrm{CO}}$,
and considerable extinction ($\tau_{\mathrm{dust}} \gtrsim 1.5$; from dust models of \citet{laor93} extrapolated to our
wavelengths) of the S(1) line, we find it to be consistent with our mass range.

\begin{table}[t]
\caption{Model results and their uncertainties for the warm gas}
\label{besttable}
\centering
\begin{tabular}{l c l}
\hline\hline
Quantity  &  Most Probable Value  & range$^\dagger$ \\ 
\hline
$T_{\textrm{kin}}$ (K)  & 545 & 350 -- 825  \\
$\mathrm{Log}_{10}\,n({\mathrm{H}_2})$ (cm$^{-3}$) & 3.7   & 3.0 -- 4.1 \\
Log$_{10} N(^{12}$CO) (cm$^{-2}$) & 19.0 & 18.5 -- 19.8\\
$\mathrm{Log}_{10}\,\Phi_{A}N(^{12}$CO)$^\ddagger$ (cm$^{-2})$ & 17.4 & 17.2 -- 17.9 \\
$N(^{12}$CO)/$N(^{13}$CO)   & $ 20 $ & $15 - 37$ \\
Log$_{10} P$ (K cm$^{-3}$) & 6.4 & 5.8 -- 6.7\\
$M_{\mathrm{gas}}$ ($\times 10^{7}\,\mathrm{M}_{\sun}$) & 1.2  &  0.7 -- 3.6  \\ 
\hline
\multicolumn{3}{l}{$^\dagger$ Ranges are for 95\% confidence intervals}\\
\multicolumn{3}{l}{$^\ddagger$ Beam averaged column density where $\Phi_{\mathrm{A}}$ is an area filling factor}
\end{tabular}
\end{table}

%\begin{table}
%\begin{minipage}[t]{\columnwidth}
%\caption{Model results and their uncertainties for the warm gas.}
%\label{besttable}
%\centering
%\renewcommand{\footnoterule}{}  % to avoid a line before footnotes
%\begin{tabular}{l c l}
%\hline\hline
%Quantity  &  Most Probable Value  & range\footnote{Ranges are for 95\% confidence intervals} \\ 
%\hline
%$T_{\textrm{kin}}$ (K)  & 545 & 350 -- 825  \\
%$\mathrm{Log}_{10}\,n({\mathrm{H}_2})$ (cm$^{-3}$) & 3.7   & 3.0 -- 4.1 \\
%Log$_{10} N(^{12}$CO) (cm$^{-2}$) & 19.0 & 18.5 -- 19.8\\
%$\mathrm{Log}_{10}\,\Phi_{A}N(^{12}$CO)\footnote{Beam averaged column density where $\Phi_{\mathrm{A}}$ is an area filling factor} (cm$^{-2})$ & 17.4 & 17.2 -- 17.9 \\
%$N(^{12}$CO)/$N(^{13}$CO)   & $ 20 $ & $15 - 37$ \\
%Log$_{10} P$ (K cm$^{-3}$) & 6.4 & 5.8 -- 6.7\\
%$M_{\mathrm{gas}}$ ($\times 10^{7}\,\mathrm{M}_{\sun}$) & 1.2  &  0.7 -- 3.6  \\ 
%\hline
%\end{tabular}
%\end{minipage}
%\end{table}

Our inferred thermal pressure ($ 3 \times 10^{6}$ K cm$^{-3}$) is comparable to both that of the M82 atomic gas as
probed by the \ion{C}{II} and \ion{O}{I} transitions \citep{kaufman99,lord95}, and the UV-shielded dense gas
\citep{naylor10}, although this does not imply pressure equilibrium between the phases. 
Our warm-component mass is also similar to the $9 \times 10^{6}$ \msun\ inferred from the photodissociation region (PDR) analysis based on the atomic gas lines \citep{kaufman99}.  However, the CO emission in the warm molecular gas likely does not arise from PDRs. This is because we measure CO transitions that are much more luminous than predicted by the PDR models. These models require an extremely high density PDR (n $> 10^{5}$) to reproduce the $J$ = 6--5 and 7--6 intensities, a condition which is clearly ruled out by the atomic lines and their ratio to the far-IR flux.

At a temperature of about 500~K, H$_{2}$ will be the dominant coolant compared to CO. This is evident from
the observed H$_{2}$ line luminosities, and agrees with the model predictions \citep{neufeld95, lebourlot95}. 
The models predict H$_{2}$ cooling of $\sim10^{-22.6}$~ergs~s$^{-1}$ per molecule for the temperature and density of SPIRE-observed warm gas. This implies a total molecular gas cooling of about 2.6 L$_{\sun}$/\msun; i.e., $1.2 \times 10^{7}$ 
\msun\ of gas will radiate about $3 \times 10^{7} \,\mathrm{L}_{\sun}$ in H$_{2}$ lines, in good agreement with the
value derived from \textit{ISO} and \textit{Spitzer} data.
% measurements. 

What is the heating source of this warm molecular gas?  Hard X-rays from an AGN  have the potential
for heating molecular gas in an XDR \citep{maloney96}, but there is no strong evidence for an AGN in \object{M82}
\citep{strickland07}. Moreover, with a strong XDR component, such as seen in \object{Mrk231} \citep{vanderwerf10},
the SLED becomes flat at high $J$ instead of decreasing as in \object{M82}.
Another scenario is heating via the enhanced cosmic ray density generated by the high supernova rate in the nuclear starburst \citep{suchkov93}.  
With a cosmic ray energy deposition rate of 3.5--$12 \times 10^{-16}$~eV~s$^{-1}$ per H$_{2}$ molecule in the Galaxy
\citep{goldsmith78,vandishoeck86}, the energy input per mass in M82 is 0.09 to 0.3 L$_{\sun}$/M$_{\sun}$, 
too low to match the observed cooling.

A second possibility is heating from the dissipation of turbulence \citep{falgarone95,maclow99,pan09}.
Using a velocity gradient of 35 km~s$^{-1}$~pc$^{-1}$ computed from our best fit RADEX model and a typical
sizescale (or Jeans length) between 0.3 to 1.6 pc in the expression for turbulent heating per unit mass
from \citet{bradford05}, we can match the observed cooling of 2.6 L$_{\sun}$/M$_{\sun}$. Our velocity
gradient is large, particularly when compared with the few km s$^{-1}$ velocity spread found
on 1~pc scales in Galactic star-forming sites, but may not be unreasonable in M82 given the powerful
stellar winds known to be present in the starburst.

%%%%%%%%%%%%%%%%%%%%%%%%%%%%%%%%%%%%%%%%%%%%%%%%%%%

\section{Conclusions}
We have presented the \textit{Herschel}-SPIRE spectroscopic observations of the
starburst galaxy \object{M82}. The spectra show a prominent CO emission-line ladder along with \ion{C}{I}
and \ion{N}{II} lines.
Radiative transfer modeling of  CO lines clearly indicates a warm gas component at $\sim$500~K in addition to the
cold ($\sim$30~K) component derived from ground-based studies. The properties of the warm
gas are strongly constrained by the high $J$ lines, observed here for the first time.
The temperature and mass of warm gas agree with the H$_2$ rotational lines observations from
\textit{Spitzer} and \textit{ISO}.
At this temperature H$_{2}$ is the dominant coolant instead of CO, and we argue that turbulence
from stellar winds and supernovae may be the dominant heating mechanism.

%%%%%%%%%%%%%%%%%%%%%%%%%%%%%%%%%%%%%%%%%%%%%%%%%%%%%%%%%%

\begin{acknowledgements}
We are grateful to P. Maloney for his 
advices on radiative transfer modeling, and to the SPIRE FTS team
for assistance with data reduction.
SPIRE has been developed by a consortium of institutes led by
Cardiff University (UK) and including Univ. Lethbridge (Canada);
NAOC (China); CEA, OAMP (France); IFSI, Univ. Padua (Italy); IAC (Spain); 
Stockholm Observatory (Sweden); Imperial College London,
RAL, UCL-MSSL, UKATC, Univ. Sussex (UK); and Caltech/JPL, IPAC,
Univ. Colorado (USA). This development has been supported by
national funding agencies: CSA (Canada); NAOC (China); CEA,
CNES, CNRS (France); ASI (Italy); MCINN (Spain); SNSB (Sweden); 
STFC (UK); and NASA (USA).
Additional funding support for some instrument
activities has been provided by ESA. 
\end{acknowledgements}

%%%%%%%%%%%%%%%%%%%%%%%%%%%%%%%%%%%%%%%%%%%%%%%%%%%%%%%%%%


\begin{thebibliography}{}


\bibitem[\protect\citeauthoryear{Beir\~ao et al.}{2008}]{beirao08}
Beir\~ao, P., Brandl, B. R., Appleton, P. N., et al. 2008, ApJ, 676, 304

\bibitem[\protect\citeauthoryear{Bradford et al.}{2005}]{bradford05}
Bradford, C. M., Stacey, G. J., Nikola, T., et al. 2005, ApJ, 623, 866

\bibitem[\protect\citeauthoryear{Colbert et al.}{1999}]{colbert99}
Colbert, J. W., Malkan, M. A., Clegg, P. E., et al. 1999, ApJ, 511, 721

\bibitem[\protect\citeauthoryear{Falgarone \& Puget}{1995}]{falgarone95}
Falgarone, E., Puget, J.-L. 1995, A\&A, 293, 840

\bibitem[\protect\citeauthoryear{F\"orster Schreiber et al.}{2003}]{forster03}
F\"orster Schreiber, N. M., Genzel, R., Lutz, D., \& Sternberg, A. 2003, ApJ, 599, 193

\bibitem[\protect\citeauthoryear{Goldsmith \& Langer}{1978}]{goldsmith78}
Goldsmith, P. F., Langer, W. D. 1978, ApJ, 222, 881

\bibitem[\protect\citeauthoryear{Griffin et al.}{2010}]{griffin10}
Griffin, M. et al. 2010, A\&A, this volume

\bibitem[\protect\citeauthoryear{Kaufman et al.}{1999}]{kaufman99}
Kaufman, M. J., Wolfire, M. G., Hollenbach, D J., \& Luhman, M. L. 1999, ApJ, 527, 795

\bibitem[\protect\citeauthoryear{Kaufman, Wolfire \& Hollenbach}{Kaufman et al.}{2006}]{kaufman06}
Kaufman, M. J., Wolfire, M. G., \& Hollenbach, D. J. 2006, ApJ, 644, 283

\bibitem[\protect\citeauthoryear{Mac Low}{1999}]{maclow99}
Mac Low, M.-M. 1999, ApJ, 524, 169

\bibitem[\protect\citeauthoryear{Maloney, Hollenbach, \& Tielens}{Maloney et al.}{1996}]{maloney96}
Maloney, P. R., Hollenbach, D. J., \& Tielens, A. G. G. M. 1996, ApJ, 466, 561

\bibitem[\protect\citeauthoryear{Naylor et al.}{2010}]{naylor10}
Naylor B. J., Bradford, C. M., Aguirre, J. E., et al. 2010, ApJ, submitted

\bibitem[\protect\citeauthoryear{Naylor \& Tahic}{2007}]{naylor07}
Naylor D. A., \& Tahic M. K., 2007, J. Opt. Soc. Am. A, 24, 3644


\bibitem[\protect\citeauthoryear{Neufeld, Lepp, \& Melnick}{Neufeld et al.}{1995}]{neufeld95}
Neufeld, D. A., Lepp, S., Melnick, G. J. 1995, ApJS, 100, 132

\bibitem[\protect\citeauthoryear{Laor \& Draine}{1993}]{laor93}
Laor, A., \& Draine, B. T. 1993, ApJ, 402, 441

\bibitem[\protect\citeauthoryear{Le Bourlot, Pineau des For\^ets \& Flower}{Le Bourlot et al.}{1998}]{lebourlot95}
Le Bourlot, J., Pineau des For\^ets, G., \& Flower, D. R. 1999, MNRAS, 305, 802

\bibitem[\protect\citeauthoryear{Lord et al.}{1995}]{lord95}
Lord, S. D., Hollenbach, D. J., Colgan, S. W. J. et al. 1995, ASPC, 73, 151

\bibitem[\protect\citeauthoryear{Pan \& Padoan}{2009}]{pan09}
Pan, L., \& Padoan, P. 2009, ApJ, 692, 594

\bibitem[\protect\citeauthoryear{Petitpas \& Wilson}{2000}]{petitpas00}
Petitpas, G. R., \& Wilson, C. D. 2000, ApJ, 538, L117

\bibitem[\protect\citeauthoryear{Pilbratt et al.}{2010}]{pilbratt10}
Pilbratt, G. et al. 2010, A\&A, this volume 

\bibitem[\protect\citeauthoryear{Rigopoulou et al.}{2002}]{rigopoulou02}
Rigopoulou, D., Kunze, D., Lutz, D., Genzel, R., \& Moorwood, A. F. M. 2002, A\&A, 389, 374

\bibitem[\protect\citeauthoryear{Roussel et al.}{2010}]{roussel10}
Roussel, H. et al. 2010, A\&A, this volume

\bibitem[\protect\citeauthoryear{Sakai \& Madore}{1999}]{sakai99}
Sakai, S. \& Madore, B. F. 1999, ApJ, 526, 599

\bibitem[\protect\citeauthoryear{Sanders et al.}{2003}]{sanders03}
Sanders, D. B., Mazzarella, J. M., Kim, D.-C.,
Surace, J. A., \& Soifer, B. T. 2003, AJ, 126, 1607

\bibitem[\protect\citeauthoryear{Seaquist, Lee \& Moriarty-Schieven}{Seaquist et al.}{2006}]{seaquist06}
Seaquist, E. R., Lee, S. W., \& Moriarty-Schieven, G. H. 2006, ApJ, 638, 148

\bibitem[\protect\citeauthoryear{Shen \& Lo}{1995}]{shen95}
Shen, J. \& Lo, K. Y., 1995, ApJ, 445, L99

\bibitem[\protect\citeauthoryear{Strickland \& Heckman}{2007}]{strickland07}
Strickland, D. K., \& Heckman, T. M. 2007, ApJ, 658, 258

\bibitem[\protect\citeauthoryear{Suchkov, Allen, \& Heckman}{Suchkov et al.}{1993}]{suchkov93}
Suchkov, A., Allen, R. J., \& Heckman, T. M. 1993, ApJ, 413, 542

\bibitem[\protect\citeauthoryear{Swinyard et al.}{2010}]{swinyard10}
Swinyard, B. G., Ade, P., Baluteau, J-P. et al. 2010, A\&A, this volume

\bibitem[\protect\citeauthoryear{Tilanus et al.}{1991}]{tilanus91}
Tilanus, R. P. J., Tacconi, L. J., Sutton, E. C.,
Zhou, S., Sanders, D. B., Wynn-Williams, C. G., Lo, K. Y., \& Stephens,
S. A., 1991, ApJ, 376, 500

\bibitem[\protect\citeauthoryear{van der Tak et al.}{2007}]{vandertak07}
van der Tak, F. F. S., Black, J. H., Sch\"oier, F. L., Jansen, D. J., \& van Dishoeck, E. F.
2007, A\&A, 468, 627

\bibitem[\protect\citeauthoryear{van der Werf et al.}{2010}]{vanderwerf10}
van der Werf et al. 2010, A\&A, this volume

\bibitem[\protect\citeauthoryear{van Dishoeck, \& Black}{1986}]{vandishoeck86}
van Dishoeck, E. F., Black, J. H. 1986, ApJS, 62, 109

\bibitem[\protect\citeauthoryear{Walter, Weiss \& Scoville}{Walter et al.}{2002}]{walter02}
Walter, F, Weiss, A., \& Scoville, N. 2002, ApJ, 580, L21

\bibitem[\protect\citeauthoryear{Ward et al.}{2003}]{ward03}
Ward, J. S., Zmuidzinas, J., Harris, A. I., \&
Isaak, K. G. 2003, ApJ, 587, 171 (W03)

\bibitem[\protect\citeauthoryear{Weiss, Walter \& Scoville}{Weiss et al.}{2005}]{weiss05}
Weiss, A., Walter, F., \& Scoville N. Z. 2005, A\&A, 438, 533

\bibitem[\protect\citeauthoryear{Wild et al.}{1992}]{wild92}
Wild, W., Harris, A. I., Eckart, A., et al. 1992, A\&A, 265, 447 

\bibitem[\protect\citeauthoryear{Yao}{2009}]{yao09}
Yao, L. 2009, ApJ, 705, 766

\bibitem[\protect\citeauthoryear{Yun, Ho, \& Lo}{Yun et al.}{1993}]{yun93}
Yun, M. S., Ho, P. T. P., \& Lo, K. Y., 1993, ApJ, 411, L17

\end{thebibliography}
\end{document}